\documentclass[reprint,english,notitlepage,aps,pra]{revtex4-2}
\usepackage[utf8]{inputenc}
\usepackage{amsmath}
\usepackage{amssymb}
\usepackage{graphicx}
\usepackage[unicode,colorlinks,citecolor=Blue3,linkcolor=Blue3,bookmarks=true]{hyperref}

\makeatletter

\usepackage{graphicx}
\usepackage[x11names]{xcolor}
\usepackage{newtxtext}
\usepackage{newtxmath}
\usepackage{verbatim}
\usepackage[unicode,colorlinks,citecolor=Blue3,linkcolor=Blue3,bookmarks=true]{hyperref}

\newcommand{\eg}{{\it e.g.}}
\newcommand{\etc}{{\it etc.}}

\newcommand{\sign}{\mathop{\rm sign}\nolimits}
\renewcommand{\Re}{\mathop{\rm Re}\nolimits}
\renewcommand{\Im}{\mathop{\rm Im}\nolimits}
\newcommand{\intinfty}{\displaystyle\int_{-\infty}^{\infty}\!}

\newcommand{\F}{\mathcal{F}}
\newcommand{\K}{\mathcal{K}}
\newcommand{\mean}[1]{\langle#1\rangle}

\makeatother

\begin{document}

\title{Quantum limits for stationary force sensing}

\author{Farid Ya.\ Khalili}
\email{farit.khalili@gmail.com}
\affiliation{Russian Quantum Center, Moscow, Russia}
\affiliation{NUST ``MISiS'', Leninskiy Prospekt~4, 119049 Moscow, Russia.}

\author{Emil Zeuthen}
\email{zeuthen@nbi.ku.dk}
\affiliation{Niels Bohr Institute, University of Copenhagen, DK-2100 Copenhagen, Denmark}


\begin{abstract}

State-of-the-art sensors of force, motion, and magnetic fields have reached the sensitivity where the quantum noise of the meter is significant or even dominant. In particular, the sensitivity of the best optomechanical devices has reached the Standard Quantum Limit (SQL), which directly follows from the Heisenberg uncertainty relation and corresponds to balancing the measurement imprecision and the perturbation of the probe by the quantum back action of the meter.

The SQL is not truly fundamental and several methods for its overcoming have been proposed and demonstrated. At the same time, two quantum sensitivity constraints which are more fundamental are known. The first limit arises from the finiteness of the probing strength (in the case of optical interferometers --- of the circulating optical power) and is known as the Energetic Quantum Limit or, in a more general context, as the Quantum Cram\'{e}r-Rao Bound (QCRB).
The second limit arises from the dissipative dynamics of the probe, which prevents full efficacy of the quantum back action evasion techniques developed for overcoming the SQL. No particular name has been assigned to this limit; we propose the term \textit{Dissipative Quantum Limit} (DQL) for it.

Here we develop a unified theory of these two fundamental limits by deriving the general sensitivity constraint from which they follow as particular cases. Our analysis reveals a phase transition occurring at the boundary between the QCRB-dominated and the DQL regimes, manifested by the discontinuous derivatives of the optimal spectral densities of the meter field quantum noise. This leads to the counter-intuitive (but favorable) finding that quantum-limited sensitivity can be achieved with certain lossy meter systems. Finally, we show that the DQL originates from the non-autocommutativity of the internal thermal noise of the probe and that it can be overcome in non-stationary measurements.

\end{abstract}

\maketitle

\section{Introduction}\label{sec:intro}
The improvement of sensors for minuscule signals, \eg, forces or magnetic fields, is increasingly being limited by the constraints imposed by quantum mechanics.
In the field of quantum optomechanics, laser interferometric gravitational-wave
(GW) detectors~\cite{PRL_116_131103_2016} have reached a sensitivity which allows regular observation of GW signals generated by collisions of black holes
and neutron stars~\cite{PhysRevX.9.031040}. This sensitivity
is already close to the Standard Quantum Limit (SQL), at which the measurement imprecision and the mechanical
perturbation due to the quantum back action balance~\cite{67a1eBr,74a1eBrVo,Caves_RMP_52_341_1980}.
Several methods for overcoming the SQL suitable for future GW detectors
were proposed and are under development now, see, \eg, the review articles~\cite{12a1DaKh, 19a1DaKhMi}.

In parallel, much smaller (table-top) parametric optomechanical
and microwave-mechanical devices working at quantum sensitivity level
have been developed in several laboratories, see, \eg, the review
papers~\cite{Aspelmeyer_RMP_86_1391_2014,16a1DaKh}. Harmonic oscillators with
typical eigenfrequencies ranging from hundreds of kHz to GHz
were used as mechanical objects in these experiments, in contrast to
the almost freely suspended test masses of the GW detectors. Sensitivities
close to the SQL were already demonstrated with these devices~\cite{Teufel2009, Anetsberger_NPhys_5_909_2009, 11a1WeFrKaYaGoMuDaKhDaSc, Purdy_Scence_339_801_2013} and first attempts to overcome it were made~\cite{Wollman_Science_349_952_2015, Ockeloen-Korppi_PRL_117_140401_2016, Kampel_PRX_7_021008_2017, Moeller_Nature_547_191_2017, Mason_NPhys_15_745_2019}.

Another example of quantum-limited sensors are the state-of-the-art magnetometors based on atomic spin oscillators~\cite{Wasilewski_PRL_104_133601_2010}.

\begin{figure}[ht]
\includegraphics[scale=1]{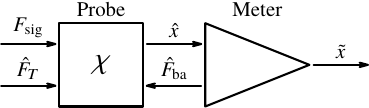} \caption{Generic linear force sensor consisting of probe and meter subsystems. The probe is subjected to an unknown classical signal force $F_{\text{sig}}$, a thermal force $\hat{F}_T$ due to dissipation, and a back-action force $\hat{F}_{\text{ba}}$ from the meter. $\chi$ is the ``bare'' probe susceptibility. $\tilde{x}$ is the meter output referenced to the probe position.}
\label{fig:linear}
\end{figure}

The distinctive feature of these experiments is the extremely small values of the signal and noise displacements --- much smaller than the corresponding characteristic scales of nonlinearities (like the probing light wavelength in the optical interferometers). This feature greatly simplifies the theoretical analysis of these systems, allowing one to use  the unified model of the \textit{linear probe system}  shown in Fig.~\ref{fig:linear}.
Here, a canonical position coordinate $\hat{x}$ of the {\it linear probe} is measured by the {\it linear meter},
whose output signal is equal to the sum of (amplified) $\hat{x}$ and the imprecision (measurement) noise $\hat{x}_{{\rm fl}}$. In return, the meter applies
the back action force $\hat{F}_{{\rm ba}}$ to the probe, consisting
of the regular dynamic part (proportional to $\hat{x}$) and the back action
noise $\hat{F}_{{\rm fl}}$.

This interaction can be described by two linear equation, which, in the particular case of a stationary meter, can be stated in the Fourier picture as
\begin{subequations}\label{meter_Omega}
  \begin{gather}
    \tilde{x}(\Omega) = \hat{x}_{{\rm fl}}(\Omega) + \hat{x}(\Omega) \,,
      \label{tilde_x} \\
    \hat{F}_{{\rm ba}}(\Omega) = \hat{F}_{{\rm fl}}(\Omega) - K(\Omega)\hat{x}(\Omega)\,,
      \label{F_ba}
  \end{gather}
\end{subequations}
where $\tilde{x}$ is the meter output referenced to the position signal, and $K$ is the dynamic susceptibility modification introduced by the meter into the probe dynamics. In the case of optical interferometers, this effect is known as the {\it optical spring}~\cite{99a1BrKh, Buonanno2002, 12a1DaKh}, with its real part $\Re K$ being the {\it optical rigidity} and its imaginary part $\Im K$ being the {\it optical damping}. The sign convention for $K$ is here chosen according to the standard definition of the Hooke factor.

If the goal of the setup in Fig.~\ref{fig:linear} is the detection of a classical signal force $F_{{\rm sig}}$ acting on the probe, then Eqs.~\eqref{meter_Omega} should be supplemented by a third one describing the probe dynamics:
\begin{equation}\label{probe_Omega}
  \chi^{-1}(\Omega)\hat{x}(\Omega)
  = F_{{\rm sig}}(\Omega) + \hat{F}_{\rm ba}(\Omega) + \hat{F}_T(\Omega) + \ldots,
\end{equation}
where $\chi$ is the probe susceptibility function, $\hat{F}_T$ is its thermal noise, and ``$\dots$'' stands for possible other forces acting on the probe. The spectral density of the thermal force $\hat{F}_T$ is, in accordance with the Fluctuation-Dissipation Theorem (FDT)~\cite{Callen1951},
\begin{multline}
  S_{\rm FDT}(\Omega)
  = \hbar|\Im\chi^{-1}(\Omega)|\coth\left(\frac{\hbar\Omega}{2k_{B}T}\right) \\
  \ge \hbar|\Im\chi^{-1}(\Omega)|, \label{FDT}
\end{multline}
where $k_{B}$ is the Boltzmann constant and $T$ is the temperature [in the case of a non-equilibrium bath, \eg, a probe system consisting of a mechanical mode damped by an auxiliary optical mode, Eq.~\eqref{FDT} generalizes straightforwardly in terms of an effective, frequency-dependent temperature, $T\rightarrow T_{\rm eff}(\Omega)$].

The first analysis of the quantum limitations of the linear probe system was done by Giffard in 1976~\cite{Giffard_PRD_14_2478_1976}, using the earlier work of Heffner~\cite{Heffner_ProcIRE_50_1604_1962}. It was shown that if the meter noise sources $\hat{x}_{\rm fl}$ and $\hat{F}_{\rm fl}$ are stationary, then their spectral densities satisfy the uncertainty relation
\begin{equation}\label{SxSF_simple}
  S_{xx}(\Omega)S_{FF}(\Omega) \ge \frac{\hbar^2}{4} \,.
\end{equation}
It is easy to show that if, in addition, these noise sources are mutually uncorrelated, $S_{xF}=0$, and the dynamic back action is absent, $K=0$, then the force sensitivity of the linear probe is limited by the SQL,
\begin{equation}
  S_{{\rm SQL}}(\Omega)=\hbar|\chi^{-1}(\Omega)|\,.\label{SQL}
\end{equation}
Here $S_{{\rm SQL}}$ is the spectral density of the equivalent force noise
and $\Omega$ is the observation (running) frequency (in Ref.\,\cite{Giffard_PRD_14_2478_1976}, a particular case of this limit for the harmonic probe oscillator and a narrow-band signal force was obtained).

This limit can also be obtained in a different (but physically equivalent)
way, namely as a consequence of non-commutativity of the probe position
operator $\hat{x}(t)$ at different moments of time (non-autocommutativity).
This approach was used in the pioneering works \cite{67a1eBr, 74a1eBrVo, Caves_RMP_52_341_1980}
where the concept of the SQL was first formulated. It was assumed
in these early works that the SQL can be evaded only using more sophisticated
Quantum-non-Demolition (QND) measurements of some autocommuting observable \cite{78a1eBrKhVo, Caves_RMP_52_341_1980}.

It was realized later by Unruh~\cite{Unruh1982} that if the goal is not the measurement of the probe position, but detection of an external classical action on
the probe, then the SQL can be evaded using position measurements with suitably cross-correlated measurement noise and back action noise. It was also shown in Ref.~\cite{Unruh1982} that in interferometric position meters, this cross-correlation can
be created by injection of squeezed light with the appropriate squeezing angle into the interferometer. A practical method for the generation of frequency-dependent squeezed light based on additional so-called filter cavities was proposed almost 20 years later in Ref.~\cite{02a1KiLeMaThVy}.

The general form of the uncertainty relation~\eqref{SxSF_simple}, which takes into account explicitly this cross-correlation, as well as the dynamic back action of the meter, was derived in Ref.~\cite{82a2eKhVo},
\begin{equation}\label{eq:Heisenberg-general}
  S_{xx}(\Omega)S_{FF}(\Omega)-|S_{xF}(\Omega)|^{2} \ge \hbar|\sigma(\Omega)|
  + \frac{\hbar^{2}}{4} \,,
\end{equation}
where $S_{xF}$ is the cross-correlation spectral density of $\hat{x}_{\rm fl}$ and $\hat{F}_{\rm fl}$ and
\begin{equation}\label{sigma}
  \sigma(\Omega) \equiv \Im\bigl\{K(\Omega)S_{xx}(\Omega) + S_{xF}^*(\Omega)\bigr\} .
\end{equation}
Note that while the left-hand side of this equation has the standard Schr\"{o}dinger-Robertson form~\cite{Schroedinger_PPAS_19_296_1930}, on the right-hand side an additional term $\hbar|\sigma|$ appears which intermixes the imaginary parts of the cross-correlation $\Im S_{xF}$ and the meter dynamic back action $\Im K$.

It was shown in Ref.~\cite{Miao_PRA_95_012103_2017} that for any bosonic system in a pure Gaussian quantum state (i) the two terms in Eq.~\eqref{sigma} cancel each other, giving $\sigma=0$, and (ii) the remaining Schr\"{o}dinger-Robertson inequality~\eqref{eq:Heisenberg-general} is saturated.
Direct calculation of the quantum noise in the case of ideal lossless optical  interferometers gives the same result, see, \eg, Ref.~\cite{12a1DaKh}. 

Constraints for quantum-limited stationary sensing were discussed also in Ref.~\cite{Clerk_RMP_82_1155_2010}, however, the analysis there was based on a meter-noise uncertainty relation which is generally weaker than Eq.~\eqref{eq:Heisenberg-general} used in the present work. In part, this stems from the fact that the uncertainty relation of Ref.~\cite{Clerk_RMP_82_1155_2010} does not account for dynamic back action damping $\Im K\neq 0$.
For these reasons, the present work reaches conclusions that differ with Ref.~\cite{Clerk_RMP_82_1155_2010}; in particular, one of our central findings is that neither $\Im S_{xF}=0$ nor $\sigma=0$ are universal requirements for optimal quantum-limited sensitivity, thus establishing a wider class of meter correlations that permit such sensitivity.

The minimization of the linear meter quantum noise with account of the relation~\eqref{eq:Heisenberg-general} reveals two sensitivity limits that are more fundamental in character than the SQL. Both of them stem from the Heisenberg uncertainty relation, but the specific physical mechanisms are different in these two cases.

The first limit arises from the finiteness of the probing strength. A given force sensitivity imposes a necessary requirement on the coupling strength between the meter and the probe, which translates into a requirement on the magnitude of the back-action force as parametrized by its spectral density $S_{FF}$. A lower bound for the force sensitivity, $S_{\rm sum} \geq S_{\rm QCRB}$, achievable for a given $S_{FF}$ is~\cite{Miao_PRL_119_050801_2017}
\begin{equation}\label{QCRB_simple}
  S_{\rm QCRB}(\Omega) \geq \frac{\hbar^2|\chi^{-1}(\Omega) + K(\Omega)|^2}{4S_{FF}(\Omega)} \,.
\end{equation}
In interferometric position meters, $S_{FF}$ is proportional to the optical power circulating in the interferometer and therefore this limit is known as the Energetic Quantum Limit~\cite{00p1BrGoKhTh}. It was shown in Ref.~\cite{Tsang_PRL_106_090401_106} that it follows from the general Quantum Cram\'er-Rao Bound (QCRB)~\cite{HelstromBook}; we will use this latter term here.
In Ref.~\cite{Miao_PRL_119_050801_2017} it was shown that in presence of dynamical damping $\Im K\ne0$ the optimized sensitivity cannot achieve the bound~\eqref{QCRB_simple}, {\it i.e.}, it is not a {\it tight} lower bound, generally. However, for a lossless probe $\Im \chi^{-1}=0$ and quantum-limited meter correlations (Eq.~\eqref{eq:Heisenberg-general} with equality), the optimized sensitivity is at most twice the spectral density on the right-hand side of Eq.~\eqref{QCRB_simple}~\cite{Miao_PRL_119_050801_2017}. In this work we improve on these weak bounds by deriving an exact expression for the optimized sensitivity for arbitrary $K$ and $\Im \chi^{-1}$.

We now turn to the second limit, which arises from the dissipative dynamics of the probe system. It was shown in Refs.~\cite{87a1eKh} (for a general linear measurement) and~\cite{JaekelReynaud1990} (for the particular case of an interferometric measurement) that the stationary cross-correlation of the measurement noise and the back-action noise, proposed in Ref.~\cite{Unruh1982}, can only compensate for the real part of the inverse probe response function $\Re \chi^{-1}$, leaving a noise contribution proportional to the imaginary (dissipative) part $\Im \chi^{-1}$,
\begin{equation}
  S_{{\rm DQL}}(\Omega)=\hbar|\Im\chi^{-1}(\Omega)|\,,\label{DQL}
\end{equation}
which can be viewed as a refined form of the (non-fundamental) SQL~\eqref{SQL}.
No particular name was proposed for this limit in Refs.~\cite{87a1eKh, JaekelReynaud1990}.
In the experimental work~\cite{Kampel_PRX_7_021008_2017}, where a sensitivity close to this limit was achieved, it was simply referred to as the ``Quantum Limit''. Here we will use the term \textit{Dissipative Quantum Limit} (DQL).
Note that this limit, which describes the minimal possible value of the \textit{meter sum quantum noise}, should not be confused with the \textit{thermal noise of the probe itself}, Eq.~\eqref{FDT}, even though the latter equals the DQL~(\ref{DQL}) when $T\to0$, for reasons to be explained in due course.
In Refs.\,\cite{87a1eKh, JaekelReynaud1990}, the DQL was derived by means of straightforward optimization of the linear position meter quantum noise, whereby the physical origin of the DQL remained obscure and the question of whether it is possible to evade it (as is the case for the SQL) is unanswered.

The goal of this paper is to develop a general, unified theory of the QRCB and the DQL, valid for all stationary linear meters, and to establish the QCRB and the DQL as  corresponding to different regimes of a general quantum noise minimization. The paper is structured as follows. In Sec.~\ref{sec:Linear-meas-theory} we review the basic principles of linear measurement theory developed in Ref.~\cite{82a2eKhVo} and introduce a convenient gauge transformation of the back action noise $\hat{F}_{\rm fl}$, generalizing the concept of effective back action $\hat{\F}_{\text{fl}}$ introduced in Ref.~\cite{Buonanno2002}. In Sec.~\ref{sec:Sens-limit} we use the formalism of Sec.~\ref{sec:Linear-meas-theory} to derive the general forms of the quantum limits for stationary force sensing and the conditions to attain them. In particular, this analysis reveals a phase-transition-like feature in the optimized quantum noise occurring at the boundary between the QCRB and the DQL regimes. In Sec.~\ref{sec:spin_system}, we discuss in more detail the factor \eqref{sigma} and provide an example of a meter with $\sigma\ne0$. In Sec.~\ref{sec:DQL} we analyze the DQL using the time-domain picture, which allows us to reveal the physical origin of the DQL; in turn this suggests a principle for overcoming this limit. In Sec.~\ref{sec:conclusion} we recapitulate the main results of this paper.

\section{Linear measurement theory\label{sec:Linear-meas-theory}}

\subsection{Conventions}\label{sec:conv}

In this paper we use the symmetrized correlation functions defined for any fluctuating operators $\hat{p}$ and $\hat{q}$ as
\begin{equation}\label{conv_B}
  B_{pq}(t,t') = \frac{\mean{\hat{p}(t)\hat{q}(t') + \hat{q}(t')\hat{p}(t)}}{2} \,.
\end{equation}
To describe stationary noise sources, we use the corresponding symmetrized, double-sided spectral densities, defined via
\begin{equation}\label{stat_B}
  B_{pq}(t,t') = B_{pq}(t-t',0)
  = \intinfty S_{pq}(\Omega)e^{-i\Omega(t-t')}\,\frac{d\Omega}{2\pi} \,.
\end{equation}

\subsection{Basic principles}\label{linear_t}

We start with the first component of the scheme shown in Fig.\,\ref{fig:linear},
the probe. Its linearity allows us to describe the dynamics of its position
$\hat{x}$ by the susceptibility function $\chi(t,t')$ as,
\begin{equation}
\hat{x}(t)=\hat{x}_{0}(t)+\intinfty\chi(t,t')\hat{F}(t')\,dt'\,,\label{probe_x_1}
\end{equation}
where $\hat{F}(t)$ is any external force acting on the probe and $\hat{x}_{0}$
is the eigenmotion of the probe in the case of $\hat{F}(t)=0$.
According to the Kubo theorem~\cite{Kubo1956}, for any linear system, $\chi(t,t')$ and the autocommutator of $\hat{x}_{0}$ can be expressed through each other,
\begin{equation}\label{probe_Kubo}
  [\hat{x}_{0}(t),\hat{x}_{0}(t')]=i\hbar\bigl\{\chi(t',t)-\chi(t,t')\bigr\}\,.
\end{equation}
The eigenmotion $\hat{x}_{0}$ can be viewed as the result of the noise force $\hat{F}_{T}$ created by the internal dissipation in the probe,
\begin{equation}
\hat{x}_{0}(t)=\intinfty\chi(t,t')\hat{F}_{T}(t')\,dt'\,.\label{probe_x_0}
\end{equation}
It follows from Eq.~(\ref{probe_Kubo}) that the autocommutator
for this force is equal to
\begin{equation}
  C_{TT}(t,t') = [\hat{F}_{T}(t),\hat{F}_{T}(t')]
    = i\hbar\bigl\{\chi^{-1}(t,t')-\chi^{-1}(t',t)\bigr\}\,,\label{probe_Kubo_F}
\end{equation}
where the inverse probe response function $\chi^{-1}$ is defined as follows:
\begin{equation}
\intinfty\chi(t,t'')\chi^{-1}(t'',t')\,dt''=\delta(t-t')\,.
\end{equation}

The second component of the scheme in Fig.~\ref{fig:linear}, the
linear position meter, can be considered in a similar way, with the
only difference being that it has two ports instead of one [compare with Eqs.\,\eqref{meter_Omega}]:
\begin{subequations}
\label{meter_1}
\begin{gather}
\tilde{x}(t)=\hat{x}_{{\rm fl}}(t) + \hat{x}(t) \,, \label{meter_1_y}\\
\hat{F}_{{\rm ba}}(t)=\hat{F}_{{\rm fl}}(t)-\intinfty K(t,t')\hat{x}(t')\,dt'\,;
  \label{meter_F}
\end{gather}
\end{subequations}
$\hat{x}_{\text{fl}}$ is the measurement imprecision noise (referenced to the position  $\hat{x}$) and $\hat{F}_{\text{fl}}$ is the stochastic (quantum) back action.
The essential purpose of the meter system is to produce a definite measurement result as represented by the state of an essentially classical object, \eg, bits in a classical computer. To derive meaningful quantum noise limits, our analysis must encompass the full measurement chain from the quantum probe to the classical measurement outcome, to ensure that (in principle) no additional amplification noise must be accounted for.
A necessary and sufficient condition for the meter to accomplish this is that of {\it simultaneous measurability} $[\tilde{x}(t),\tilde{x}(t')]=0$ for all $t,t'$~\cite{92BookBrKh, Buonanno2002}; the Kubo theorem can then be applied to establish the commutators
\begin{subequations}\label{meter_Kubo}
\begin{gather}
  C_{xx}(t,t') = [\hat{x}_{{\rm fl}}(t),\hat{x}_{{\rm fl}}(t')] = 0 \,, \label{eq:x-fl_comm}\\
  C_{FF}(t,t') = [\hat{F}_{{\rm fl}}(t),\hat{F}_{{\rm fl}}(t')]
    = i\hbar\bigl\{ K(t,t')-K(t',t)\bigr\}\,,\label{eq:F-fl_comm} \\
  C_{xF}(t,t') = [\hat{x}_{{\rm fl}}(t),\hat{F}_{{\rm fl}}(t')] = -i\hbar\lim_{\theta\to+0}\delta(t-t'-\theta)\,,
  \label{meterKuboCxF}
\end{gather}
\end{subequations}
which are necessarily $c$-numbers for a linear system. The particular form of the $\delta$ function in Eq.~(\ref{meterKuboCxF}) ensures a finite (forward) transfer function (from probe to meter output) while the reverse transfer function (which would map external variables coupled to the meter output port to the probe input) vanishes~\footnote{{A slightly more general starting point than the one adopted here is sometimes used in the literature. It involves the {\it unnormalized} output of the meter $Z$, whose signal part is proportional to the forward transfer function $\chi_{ZF}$. In the Fourier-domain analysis of stationary systems, $\chi_{ZF}$ can be straightforwardly absorbed in the meter output variable $\tilde{x} \equiv Z/\chi_{ZF}$, thereby recovering the formalism used in this work.}}.

The non-autocommutativity of the quantum back action operator $\hat{F}_{\text{fl}}$ arises from the time-antisymmetric (dissipative) part of the dynamic back action factor $K(t,t')$. For instance, this is manifested in cavity-optomechanical systems with a detuned drive, in which amplitude-to-phase interconversion mixes non-commuting input light quadratures in forming the dynamic back-action loop.

From the simple fact that $\langle\hat{\mathcal{Q}}^{\dagger}\hat{\mathcal{Q}}\rangle\geq 0$ for any operator $\hat{\mathcal{Q}}$,
the following uncertainty relations, expressed in terms of the commutators  (\ref{probe_Kubo_F}, \ref{meter_Kubo}) and the corresponding symmetrized correlation functions \eqref{conv_B}, can be derived straightforwardly using the identity for arbitrary operators $\hat{p}$ and $\hat{q}$, $\mean{\hat{p}(t)\hat{q}(t')} = B_{pq}(t,t') + C_{pq}(t,t')/2$:
\begin{equation}
  \iint_{-\infty}^{\infty} Q^*(t)Q(t')\bigl\{B_{TT}(t,t') + C_{TT}(t,t')/2\bigr\}dtdt'
  \ge 0 \,,  \label{probe_td}
\end{equation}
\begin{multline}
  \sum_{p,q\in\{x,F\}}\iint_{-\infty}^{\infty} Q_p^*(t)Q_q(t') \\
    \times\bigl\{B_{pq}(t,t') + C_{pq}(t,t')/2\bigr\}dtdt' \ge 0 \,, \label{Heisenberg_td}
\end{multline}
where $Q$, $Q_x$, $Q_F$ are arbitrary complex functions of time, $B_{TT}$, $B_{xx}$, $B_{FF}$ are the  auto-correlation functions of $\hat{F}_T$, $\hat{x}_{{\rm fl}}$, $\hat{F}_{{\rm fl}}$, and $B_{xF}$ is the cross-correlation between $\hat{x}_{{\rm fl}}$ and $\hat{F}_{{\rm fl}}$~\cite{82a2eKhVo}. Equations~(\ref{probe_td},~\ref{Heisenberg_td}) are the time-domain quantum constraints on the symmetrized correlation functions $B_{pq}$ arising from the operator non-commutativity encoded in $C_{pq}$.

Now we can join the two subsystems together. Combining Eqs.~(\ref{probe_x_1},~\ref{meter_F}) and taking into account that the external force in
Eq.~(\ref{probe_x_1}) consists of the signal force and the meter
back action,
\begin{equation}
  \hat{F}(t) = F_{{\rm sig}}(t) + \hat{F}_{{\rm ba}}(t)\,,
\end{equation}
we obtain the following equation of motion for the probe position:
\begin{equation}
  \intinfty\chi_K^{-1}(t,t')\hat{x}(t')\,dt'
  = F_{{\rm sig}}(t)+\hat{F}_{T}(t)+\hat{F}_{{\rm fl}}(t) \,, \label{probe_x_eq}
\end{equation}
where
\begin{equation}
  \chi_K^{-1}(t,t') = \chi^{-1}(t,t') + K(t,t') \,.
\end{equation}
Therefore, the meter output referenced to the signal input of the probe, {\it i.e.}, the signal force estimate, is
\begin{equation}
\tilde{F}(t)=F_{{\rm sig}}(t)+\hat{F}_{{\rm sum}}(t)+\hat{F}_{T}(t)\,,\label{tilde_F}
\end{equation}
where
\begin{equation}
  \hat{F}_{{\rm sum}}(t)
  = \intinfty\chi_K^{-1}(t,t')\hat{x}_{{\rm fl}}(t')\,dt'+\hat{F}_{{\rm fl}}(t)
  \label{F_sum_t_raw}
\end{equation}
is the \textit{sum quantum noise} of the meter.

We remark that in the literature, the sensitivity is sometimes analyzed in position rather than force units. If the objective is to measure the position signal owing to a particular (signal) force component, this is equivalent to the sensing task considered in the present analysis, and conversion between the two conventions is achieved simply via $\hat{x}(t)=\int_{-\infty}^t\chi_{K}(t,t')\hat{F}(t')\,dt'$.

\subsection{Fourier picture}

Henceforth, we will focus mostly on the case where the probe and meter
are stationary. This implies two conditions. First, its dynamic parameters
do not change when shifted in time:
\begin{equation}\label{stat_chi}
  \chi^{-1}(t,t') = \chi^{-1}(t-t',0)
  = \intinfty\chi^{-1}(\Omega)e^{-i\Omega(t-t')}\,\frac{d\Omega}{2\pi}\,,
\end{equation}
and similarly for $K$ and $\chi_K^{-1}$, with the Fourier transforms of the (real) time-domain functions having the usual symmetry property, e.g., $\chi(-\Omega)=\chi^*(\Omega)$. Note also that the asymmetry of the back-action spectrum is linked to the dynamical damping due to Eq.~\eqref{eq:F-fl_comm},
\begin{equation}
  C_{FF}(\Omega) = -2\hbar\Im K(\Omega) \,.
\end{equation}
Second, correlation functions of the noise sources $\hat{F}_{T}$,
$\hat{x}_{{\rm fl}}$, and $\hat{F}_{{\rm fl}}$ also do not change when shifted in time, which allows us to introduce the spectral densities for them, see Eq.~\eqref{stat_B}.

The corresponding Fourier-domain form of Eq.~\eqref{F_sum_t_raw} is
\begin{equation}
  \hat{F}_{{\rm sum}}(\Omega)
  = \chi_K^{-1}(\Omega)\hat{x}_{{\rm fl}}(\Omega) + \hat{F}_{{\rm fl}}(\Omega)\,,
  \label{F_sum_Omega}
\end{equation}
and the spectral density of this noise is equal to
\begin{multline}
  S_{{\rm sum}}(\Omega) = |\chi_K^{-1}(\Omega)|^2S_{xx}(\Omega)
    + 2\Re\bigl\{\chi_K^{-1}(\Omega)S_{xF}(\Omega)\bigr\} \\
    + S_{FF}(\Omega) \,. \label{S_sum}
\end{multline}

It can be shown (see Appendix~\ref{app:meter-uncert}) that the relation~\eqref{probe_td} in the Fourier representation takes a form which resembles the Fluctuation-Dissipation Theorem~\cite{Callen1951} (albeit thermal equilibrium is not assumed here),
\begin{equation}\label{S_probe}
  S_{TT}(\Omega) \ge \hbar|\Im\chi^{-1}(\Omega)| \,,
\end{equation}
cf.\ Eq.~\eqref{FDT}, and that Eq.~\eqref{Heisenberg_td} gives the uncertainty relation~\eqref{eq:Heisenberg-general}.

\subsection{Gauge transformation of the meter noise}\label{sec:gauge}

The structure of the sum quantum noise, Eq.~\eqref{F_sum_t_raw}, allows some freedom as to what we formally identify as measurement imprecision and back-action noise, respectively:
\begin{equation}
  \hat{F}_{{\rm sum}}(t) = \intinfty\chi_\K^{-1}(t,t')\hat{x}_{{\rm fl}}(t')\,dt'
    + \hat{\F}_{{\rm fl}}(t)\,,\label{F_sum_t}
\end{equation}
where we have introduced the {\it effective} probe response function and {\it effective} back-action force,
\begin{gather}
  \chi_\K^{-1}(t,t') = \chi^{-1}(t,t')+\K(t,t')\label{eq:chi-tilde-def}\, ,\\
  \hat{\F}_{\rm fl}(t)
      = \hat{F}_{\rm fl}(t) + \intinfty\bigl\{K(t,t')-\K(t,t')\bigr\}
          \hat{x}_{\rm fl}(t')\,dt' \,,\label{eq:mathcalF-fl}
\end{gather}
in terms of an arbitrary real function $\K(t,t')$ (the {\it effective} dynamic back action factor). Accordingly we have the modified commutator,
\begin{equation}
C_{\F\F}(t,t') = [\hat{\F}_{{\rm fl}}(t),\hat{\F}_{{\rm fl}}(t')] = i\hbar\bigl\{ \K(t,t')-\K(t',t)\bigr\},\label{eq:mathcalF-fl_comm}
\end{equation}
c.f.\ Eq.~\eqref{eq:F-fl_comm}.
The above amounts to the following ``gauge'' transformation, which was first introduced in Ref.~\cite{Buonanno2002} for the particular case of $\K=0$:
\begin{subequations}\label{gauge_t}
  \begin{gather}
    \chi_K \to \chi_{\K} \,, \\
    \hat{F}_{\rm fl}(t) \to \hat{\F}_{\rm fl}(t)\,. \label{gauge_t_calF}
  \end{gather}
\end{subequations}
This invariance stems from the fact that the real physical dynamic back action and the cross-correlation of the imprecision noise and the back action noise (the virtual rigidity~\cite{12a1DaKh, 19a1ZePoKh}) affect the sum quantum noise in the same way.
An important class of gauge choices is that of time-symmetric functions $\K(t,t')=\K(t',t)$, for which $C_{\F\F}(t,t')=0$, see Eq.~\eqref{eq:mathcalF-fl_comm}. For this class, the transformation~\eqref{gauge_t_calF} amounts to excluding the non-autocommuting part of the full quantum back action $\hat{F}_{\rm fl}$ owing to the formation of the dynamic feedback loop.

\begin{figure}[th]
  \includegraphics[scale=1]{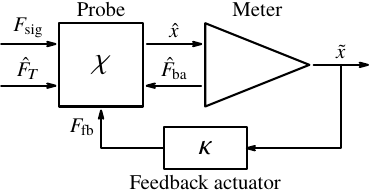}\caption{The linear force sensing scheme of Fig.~\ref{fig:linear} with an additional feedback loop based on the measurement record.}
  \label{fig:feedback}
\end{figure}

It is interesting that the formal transformation~\eqref{gauge_t} can, in principle, be implemented experimentally. Consider again the scheme of Fig.~\ref{fig:linear}, but with an added feedback loop which applies to the probe object a force proportional to the meter output signal $\tilde{x}$~\eqref{meter_1_y}, see Fig.~\ref{fig:feedback}. Such feedback has been implemented in, \eg, quantum-regime tabletop optomechanics experiments~\cite{Aspelmeyer_RMP_86_1391_2014, Sudhir_PRX_7_011001_2017, Rossi_Nature_563_53_2018}. We stress that, from a fundamental standpoint, this modification cannot improve the performance of a stationary force sensor.

With account of the feedback, the equation of motion~\eqref{probe_x_eq} takes the following form:
\begin{equation}
  \intinfty\chi_K^{-1}(t,t')\hat{x}(t')\,dt'
  = F_{{\rm sig}}(t) + \hat{F}_{T}(t) + \hat{F}_{{\rm fl}}(t) + F_{{\rm fb}}(t) \,,
\end{equation}
where
\begin{equation}
  F_{{\rm fb}}(t) = \intinfty\kappa(t,t')\tilde{x}(t')\,dt' \,,
\end{equation}
is the feedback force and $\kappa$ is the feedback factor. It can be shown using these equations that the feedback modifies the dynamic back action and  the back-action noise as follows:
\begin{subequations}\label{feedback_t}
  \begin{gather}
    K(t,t') \to K(t,t') - \kappa(t,t') \,, \label{K_res}\\
    \hat{F}_{\rm fl}(t)
      \to \hat{F}_{\rm fl}(t) + \intinfty\kappa(t,t')\hat{x}_{\rm fl}(t')\,dt' \,,
  \end{gather}
\end{subequations}
keeping the sum quantum noise $\hat{F}_{\rm sum}$ unchanged. It is easy to see that this {\it physical} modification of the meter parameters has exactly the same form as the transformation~\eqref{gauge_t} with the identification
\begin{equation}
  \K(t,t') = K(t,t') - \kappa(t,t') \,.
\end{equation}
This observation suggests that, in the present context, there is no fundamental difference between the coherent feedback loop constituted by the dynamical back action and an active measurement-based feedback.

The Fourier-domain forms of Eqs.~(\ref{F_sum_t}--\ref{eq:mathcalF-fl}) are the following:
\begin{gather}
  \hat{F}_{{\rm sum}}(\Omega)
    = \chi_\K^{-1}(\Omega)\hat{x}_{{\rm fl}}(\Omega) + \hat{\F}_{{\rm fl}}(\Omega)\,,\label{gauge_Fsum} \\
    \chi_\K^{-1}(\Omega) = \chi^{-1}(\Omega)+\K(\Omega) \,, \label{eq:chi-cK-Fourier}\\
  \hat{\F}_{\rm fl}(\Omega)
    = \hat{F}_{\rm fl}(\Omega) + \bigl\{K(\Omega)-\K(\Omega)\bigr\}
        \hat{x}_{\rm fl}(\Omega) \,. \label{gauge_Omega_calF}
\end{gather}
The spectral density of the sum quantum noise~\eqref{gauge_Fsum} is
\begin{multline}
  S_{{\rm sum}}(\Omega) = |\chi_\K^{-1}(\Omega)|^2S_{xx}(\Omega)
  + 2\Re\bigl\{\chi_\K^{-1}(\Omega)S_{x\F}(\Omega)\bigr\} \\
  + S_{\F\F}(\Omega) \,,\label{S_sum_eff}
\end{multline}
where
\begin{subequations}\label{gauge_Omega}
  \begin{multline}
    S_{\F\F}(\Omega) = |K(\Omega)-\K(\Omega)|^{2}S_{xx}(\Omega) \\
          + 2\Re\bigl(\{K(\Omega)-\K(\Omega)\}S_{xF}(\Omega)\bigr) + S_{FF}(\Omega)\,,
          \label{eq:ScalF-1}
  \end{multline}
  \begin{equation}
    S_{x\F}(\Omega) = \bigl\{K(\Omega)-\K(\Omega)\bigr\}^*S_{x}(\Omega)
      + S_{xF}(\Omega)\,, \label{eq:SxCalF}
  \end{equation}
\end{subequations}
are the spectral density of $\hat{\F}_{{\rm fl}}$ and its cross-correlation spectral density with $\hat{x}_{\rm fl}$.

Note that both the left-hand side of Eq.~\eqref{eq:Heisenberg-general} and the factor $\sigma$ are invariant under the transformation~\eqref{gauge_Omega}, which keeps the structure of the uncertainty relation unchanged,
\begin{equation}\label{eq:Heisenberg-V}
  S_{xx}(\Omega)S_{\F\F}(\Omega) - |S_{x\F}(\Omega)|^2 \ge \hbar|\sigma(\Omega)|
    + \frac{\hbar^{2}}{4} \,,
\end{equation}
where now the factor $\sigma$ has the form
\begin{equation}\label{eq:sigma-mathcal}
  \sigma(\Omega) = \Im\bigl\{\K(\Omega)S_{xx}(\Omega) + S_{x\F}^*(\Omega)\bigr\} .
\end{equation}

\section{Quantum sensitivity limits for stationary systems}\label{sec:Sens-limit}

\subsection{On the strategies of optimization}

Various approaches to optimization of the sum quantum noise spectral density~\eqref{S_sum} under the constraint of the uncertainty relation~\eqref{eq:Heisenberg-general} are possible. For any given values of $\chi$ and $K$, a triad $S_{xx}$, $S_{xF}$, and $S_{FF}$ can be found which provides the ultimate minimum of $S_{\rm sum}$. At the same time, in the derivation of the QCRB, another approach is used, namely, the conditional optimization $S_{\rm sum}$ for the given values of $\chi$, $K$, {\it and} $S_{FF}$. This approach emphasizes the fact that $S_{FF}$ is proportional to the measurement strength, which is a physical resource that could be limited by experimental constraints.

The gauge transformation of Sec.~\ref{sec:gauge} creates an additional degree of freedom for this optimization, and it could be performed for an arbitrary value of the parameter $\K$ with a result which, in the ``effective'' notations $\F$ and $\K$, has exactly the same form for all values of $\K$, including for $\K=K$. A quite strong conclusion follows from this invariance. Evidently, the ultimate minimum for a freely tunable probe strength $S_{FF}$ or $S_{\F\F}$ cannot depend on the arbitrary parameter $\K$. Therefore, {\it it can not depend on the physical dynamic back action $K$} as well.

The results of the conditional optimizations for a given $S_{\F\F}$ with various values of $\K$ correspond to different (depending on $\K$) trajectories in the $\{S_{xx},\,S_{FF},\,S_{xF}\}$ space which eventually converge to the same absolute minimum subspace where the sensitivity is saturated at the DQL. One of these trajectories, with $\K=K$, provides the smallest values of the physical back action noise spectral density $S_{FF}$ for given values of $S_{\rm sum}$ and is therefore typically the most interesting from a practical point of view. However, other choices of $\K$ could be of interest for particular purposes.

Here we start, in Sec.~\ref{sec:Qlimits-ident}, with the optimization for a given effective back action spectral density $S_{\F\F}$. We do so for the particular class of real (Fourier-domain) gauge functions
\begin{equation}\label{Re_K}
  \K(\Omega) \to \K'(\Omega) \in \mathbb{R}
\end{equation}
(the prime being a reminder of this restriction). The advantage of this case is that it, being mathematically very simple and transparent, allows us to obtain the DQL, and a form of QCRB (for fixed $S_{\F\F}$), as well as to identify the phase transition between them. This analysis also covers the case of optimization for a given physical back action noise spectral density $S_{FF}$ with $K=0$. Later, in Sec.~\ref{sec:Qlimits-gen}, we perform optimization for a given $S_{FF}$ in the general case of $K\ne0$.

\subsection{Identification of quantum limits and the phase transition between
them}\label{sec:Qlimits-ident}

In the case~\eqref{Re_K}, the straightforward minimization of $S_{\rm sum}~$(\ref{S_sum_eff}) under the constraint of Eq.~\eqref{eq:Heisenberg-V} with equality gives that the minimum for fixed $S_{\F\F}$ is provided by
\begin{subequations}\label{S_xS_xF_opt}
  \begin{equation}
    S_{xx}(\Omega) = \frac{1}{S_{\F\F}(\Omega)}\biggl\{
        |S_{x\F}(\Omega)|^2 + \hbar|\Im S_{x\F}(\Omega)| + \frac{\hbar^2}{4}
      \biggr\} ,
  \end{equation}
  \begin{multline}\label{S_xF_opt}
    S_{x\F}(\Omega)
      = -S_{\F\F}(\Omega)\Re\chi_{\K'}(\Omega) \\
      - i \Im\chi_{\K'}(\Omega)
      \max\{ 0,S_{\F\F}(\Omega)-S_{\text{thr\,0}}(\Omega)\}
  \end{multline}
\end{subequations}
and is equal to
\begin{equation}\label{S_sum_min0}
  S_{{\rm sum}}(\Omega)  = \begin{cases}
    S_{\rm UB\,0}(\Omega)\,, & \text{if}\
      S_{\F\F}(\Omega) < S_{\rm thr\,0}(\Omega)\,,\\[1ex]
    S_{\rm DQL}(\Omega)\,, & \text{if}\ S_{\F\F}(\Omega) \ge S_{\rm thr\,0}(\Omega) \,,
  \end{cases}
\end{equation}
where (noting that $\Im\chi^{-1}=\Im\chi_{\K'}^{-1}$)
\begin{equation}\label{S_UB_0}
  S_{\rm UB\,0}(\Omega) = \frac{\hbar|\Im\chi^{-1}(\Omega)|}{2}\biggl\{
      \frac{S_{\rm thr\,0}(\Omega)}{S_{\F\F}(\Omega)}
      + \frac{S_{\F\F}(\Omega)}{S_{\rm thr\,0}(\Omega)}
    \biggr\}
\end{equation}
is the universal bound which combines the QCRB and DQL for $S_{\F\F}(\Omega) \leq S_{\rm thr\,0}$, and
\begin{equation}\label{S_FF_th_0}
  S_{\rm thr\,0}(\Omega) =  \dfrac{\hbar|\chi_{\K'}^{-1}(\Omega)|^2}{2|\Im\chi^{-1}(\Omega)|}
  =  \dfrac{\hbar}{2|\Im\chi_{\K'}(\Omega)|}
\end{equation}
is the threshold value of the back action noise spectral density. $S_{\rm DQL}$ is given by Eq.~\eqref{DQL} and $\chi_{\K'}$ is given by Eq.~\eqref{eq:chi-cK-Fourier}, the prime being a reminder of the restriction~\eqref{Re_K}. 
We write the subscript `0' in order to distinguish the results of the present, fixed-$S_{\F\F}$ analysis from the subsequent fixed-$S_{FF}$ analysis.

It follows from these results that if $S_{\F\F}$ is smaller than the threshold value~\eqref{S_FF_th_0} (the ``weak back action'' case), then the first clause is realized in Eq.~\eqref{S_sum_min0}, with the first term in the curly braces of Eq.~\eqref{S_UB_0} dominating. In the limiting case of $S_{\F\F}\ll S_{\rm thr\,0}$, we may retain only this term, obtaining a bound that represents a form of QCRB for a given $S_{\F\F}$,
\begin{equation}\label{QCRB_calF}
  S_{\rm sum}(\Omega) = \frac{\hbar^2|\chi_{\K'}^{-1}(\Omega)|^2}{4S_{\F\F}(\Omega)}\,,
\end{equation}
compare with Eq.~\eqref{QCRB_simple}.
Note that in the special case $\K'=\Re K$ we have from Eq.~\eqref{eq:ScalF-1} that
\begin{multline}\label{S_calFF_noK}
  S_{\F\F}(\Omega) = S_{xx}(\Omega)\Im^2K(\Omega) - 2\Im S_{xF}(\Omega)\Im K(\Omega) \\
    + S_{FF}(\Omega) \,.
\end{multline}
Moreover, the various spectral densities scale with the circulating power $I_c$ in, \eg, optomechanical systems, as
\begin{equation}\label{eq:power-scaling}
  S_{xx} \propto I_c^{-1}\,,\quad S_{xF} \propto I_c^0\,,
  \quad S_{FF} \propto I_c^1\,,\quad \Im K\propto I_c^1 \,.
\end{equation}
Therefore, the spectral density~\eqref{S_calFF_noK} is proportional to $I_c$, and thus the limit~\eqref{QCRB_calF} is inversely proportional to $I_c$, similar to Eq.~\eqref{QCRB_simple}.

In the second, ``strong back action'' case of $S_{\F\F}\ge S_{\rm thr\,0}$ the ultimate sensitivity~\eqref{S_sum_min0} does not depend on $S_{\F\F}$ anymore
and is equal to the DQL~(\ref{DQL}). Note that while the sum noise spectral density~\eqref{S_sum_min0} is a continuous function of $S_{\F\F}$, as expected, its second derivative in $S_{\F\F}$, as well as the first derivative of $\Im S_{x\F}$ have discontinuities at $S_{\F\F}= S_{\rm thr\,0}$, which indicates a phase transition occurring at this point. The saturation of $S_{\text{sum}}$
at the DQL value as $S_{\F\F}$ increases occurs right at
the phase transition boundary of the two cases.

It is interesting that this phase transition exists only if $\Im S_{x\mathcal{F}}$ can take non-zero values. If $\Im S_{x\mathcal{F}}=0$, then only the first clause survives in Eq.~\eqref{S_sum_min0},
\begin{equation}\label{S_sum_simple_0}
  S_{{\rm sum}}(\Omega) = S_{\rm UB\,0}(\Omega)
\end{equation}
for all values of $S_{\F\F}$. In this case, $S_{{\rm sum}}$ attains its minimum, equal to the DQL~(\ref{DQL}), only at $S_{\F\F} = S_{\rm th\,0}$, so that increasing $S_{\F\F}$ further leads to an increase of $S_{\rm sum}$.

We remark that our expression for the DQL, Eq.~\eqref{DQL}, differs from the result of Ref.~\cite{Clerk_RMP_82_1155_2010}, which amounts to the replacement $\chi^{-1}\rightarrow \chi_{K}^{-1}$ in Eq.~\eqref{DQL} [see Eq.~\eqref{eq:chi-cK-Fourier} with $\K=K$]; this would allow $S_{\rm DQL}=0$ for suitably engineered $K$. We ascribe this disagreement to the weaker meter noise uncertainty relation employed in Ref.~\cite{Clerk_RMP_82_1155_2010}, as discussed in Sec.~\ref{sec:intro}.

Note that in the case of $\Im\K=0$ considered here, Eq.~\eqref{Re_K}, if $\Im S_{x\mathcal{F}}\ne0$, then $\sigma\ne0$ as well. According to Ref.~\cite{Miao_PRA_95_012103_2017}, this means the presence of dissipation in the meter. Thus, we have obtained the counter-intuitive result that dissipation in the meter can improve the sensitivity under certain circumstances, i.e., in the scenario where the sensitivity is not limited by the available power, $S_{\F\F}\geq S_{\rm thr\,0}$. Even though the minimized sum quantum noise $S_{\rm sum}$ saturates to the DQL value~\eqref{DQL} at $S_{\F\F}=S_{\rm thr\,0}$ (at which point $\Im S_{x\F}=0$), the possibility to have $\Im S_{x\mathcal{F}}\ne0$ while retaining quantum-limited sensitivity gives additional flexibility in tuning the meter noise spectral densities. An example of such a lossy meter system is provided in Sec.~\ref{sec:spin_system}.

\begin{figure}
\includegraphics[scale=0.9]{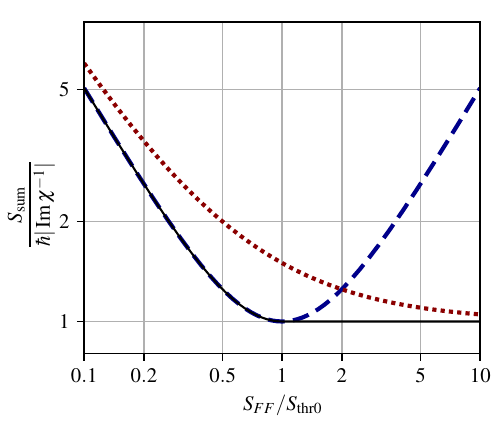}
\caption{Plots of the sum quantum
noise spectral density $S_{{\rm sum}}$ as a function of the back
action noise spectral density $S_{\F\F}$. Shown cases are the
full optimization~(\ref{S_sum_min0}) [thin solid line], the optimization under the constraint $\Im S_{x\F}=0$~(\ref{S_UB_0}) [dashed line], and the negative-mass-reference-frame measurement scheme~\cite{Hammerer_PRL_102_020501_2009, Moeller_Nature_547_191_2017}
with sub-optimally matched susceptibilities of the mechanical object
and the atomic ensemble subsystem, Eq.~(\ref{S_sum_spin}) [dotted line].}
\label{fig:plot_S_sum}
\end{figure}

The above considerations are illustrated by Fig.~\ref{fig:plot_S_sum}, in which the optimized spectral densities~(\ref{S_sum_min0}, \ref{S_sum_simple_0}) are are plotted as a function of the back-action noise spectral density $S_{\F\F}$.

\subsection{Optimization of the sum quantum noise for a given $S_{FF}$}\label{sec:Qlimits-gen}

The spectral density $S_{\F\F}$~\eqref{S_calFF_noK}, despite being qualitatively similar to the physical back action noise spectral density $S_{FF}$, depends also on $S_{xx}$ and $S_{xF}$. This means that in addition to the ``hardcore'' parameters like (in the case of optical interferometers) the optical power, bandwidth, arm length \etc, it depends also on more easily tunable factors, \eg, the homodyne angle. The optimization for a given $S_{FF}$ allows to eliminate them,
and is performed in Appendix~\ref{app:opt}. The resulting equations are more cumbersome than the ones of the previous section, but their general structure is retained. If $\sigma$ is allowed to be nonzero, as is relevant in the lossy probe case, again the same two regimes arise, with a phase transition between them:
\begin{equation}\label{S_sum_min}
  S_{{\rm sum}}(\Omega) = \begin{cases}
    S_{\rm UB}(\Omega)\,, & \text{if}\ S_{FF}(\Omega) < S_{\rm thr}(\Omega)\,,\\
    S_{\rm DQL}(\Omega) \,, & \text{if}\ S_{FF}(\Omega) \ge S_{\rm thr}(\Omega) \,,
  \end{cases}
\end{equation}
where
\begin{widetext}
\begin{equation}\label{S_UB}
  S_{\rm UB}(\Omega) = \frac{
      \hbar|\Im\chi^{-1}(\Omega)|
        \bigl\{S_{\rm thr}^2(\Omega) + S_{FF}^2(\Omega) - \hbar^2\Im^2K(\Omega)\bigr\}
    }{
      S_{\rm thr}(\Omega)S_{FF}(\Omega)
      + \sqrt{
            \bigl\{S_{\rm thr}^2(\Omega) - \hbar^2\Im^2K(\Omega)\bigr\}
            \bigl\{S_{FF}^2(\Omega) - \hbar^2\Im^2K(\Omega)\bigr\}
          }
    }
\end{equation}
\end{widetext}
and
\begin{align}
S_{\rm thr}(\Omega) &= \hbar\,\dfrac{\Re^2\chi^{-1}_{\Re K}(\Omega)+\Im^2\chi^{-1}(\Omega)+\Im^2 K(\Omega)}{2|\Im\chi^{-1}(\Omega)|}\nonumber\\
  &= \dfrac{\hbar}{2|\Im\chi^{\vphantom{1}}_{\Re K}(\Omega)|} + \dfrac{\hbar\Im^2 K(\Omega)}{2|\Im\chi^{-1}(\Omega)|}\label{S_FF_thr}
\end{align}
are the fixed-$S_{FF}$ analogues of the universal bound $S_{\rm UB\,0}$ and the threshold value $S_{\rm thr\,0}$ of the back-action noise spectral density derived in Sec.~\ref{sec:Qlimits-ident}, Eqs.~(\ref{S_UB_0},~\ref{S_FF_th_0}); $\chi_{\Re K}$ is given by Eq.~\eqref{eq:chi-cK-Fourier} with $\K=\Re K$.
Note that $S_{FF}, S_{\rm thr} \geq \hbar |\Im K|$.

Equation~\eqref{S_FF_thr} shows that whereas the optical spring $\Re K$ coherently shifts the effective probe susceptibility, the optical damping $\Im K$ gives rise to a separate, ``incoherent'' contribution to the threshold back-action strength $S_{\rm thr}$. Comparison between Eqs.~\eqref{S_FF_th_0} and \eqref{S_FF_thr} points to the particular significance of the gauge choice $\K'=\Re K$.

\begin{figure}
  \includegraphics[scale=0.9]{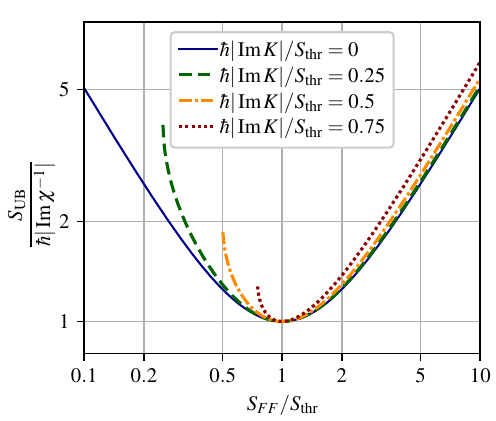} \\
  \includegraphics[scale=0.9]{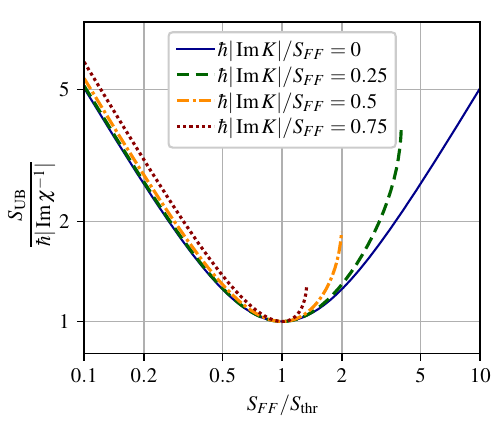} 
  \caption{Plots of $S_{\rm UB}$, Eq.~\eqref{S_UB}, as a function of $S_{FF}$ for (top) fixed $\hbar|\Im K|/S_{\rm thr}$ and (bottom) fixed $\hbar|\Im K|/S_{FF}$. Note that the plot domains are constrained by $\hbar|\Im K|\leq S_{FF},S_{\rm thr}$.}
  \label{fig:S_UB}
\end{figure}

With the constraint $\sigma=0$, again only the first clause of Eq.~\eqref{S_sum_min} survives, giving
\begin{equation}\label{S_sum_min_simple}
  S_{{\rm sum}}(\Omega) = S_{\rm UB}(\Omega)
\end{equation}
for all values of $S_{FF}$. Similarly to $S_{\rm UB\,0}$, this spectral density monotonously decreases as $S_{FF}$ increases while $S_{FF}<S_{\rm thr}$, reaching its single minimum at $S_{FF}=S_{\rm thr}$, equal to the DQL, and then starts increasing as $S_{FF}$ is increased further. This is illustrated in Fig.~\ref{fig:S_UB}~(top) where we plot $S_{\rm UB}$, Eq.~\eqref{S_UB}, in units of $S_{\rm DQL}$, as a function of $S_{FF}$ in units of $S_{\rm thr}$ for a fixed amount of dynamical damping $\Im K$.

Let us also consider the situation where the quantities $K$ and $S_{FF}$ are proportional $\hbar K(\Omega)=\alpha(\Omega)S_{FF}(\Omega)$, as is the case when, \eg, varying the circulating power in an optomechanical system while keeping the pump detuning fixed [see Eqs.~\eqref{eq:power-scaling}]. This scenario is shown in Fig.~\ref{fig:S_UB}~(bottom) for different values of $|\Im \alpha|$. Note that the rescaling factor $S_{\rm thr}$, Eq.~\eqref{S_FF_thr}, used in Fig.~\ref{fig:S_UB} depends on $\chi$ and $K$ in a manner so that the minimum at $S_{FF}=S_{\rm thr}$ cannot be achieved for arbitrary $\chi$ and $\alpha$ in the present scenario.
The above exemplifies how experimental constraints can be incorporated when applying our general analysis to a particular implementation.

It is instructive to consider the asymptotic ``pure QCRB'' limit of a lossless probe, $\Im\chi^{-1}\to0$, and compare it with the results of the work~\cite{Miao_PRL_119_050801_2017}. In this case, Eq.~\eqref{S_UB} takes the form
\begin{multline}\label{S_QCRB_gen}
  S_{\rm UB}(\Omega) \to S_{\rm QCRB}(\Omega) \\
  = \frac{\hbar^2}{2}\frac{|\chi_K^{-1}(\Omega)|^2}
      {S_{FF}(\Omega)+ \sqrt{S_{FF}^2(\Omega) - \hbar^2\Im^2K(\Omega)}} \,,
\end{multline}
where $\chi_K$ is given by Eq.~\eqref{eq:chi-cK-Fourier} with $\K=K$. 
Depending on $\Im K$, it varies between the simple form~\eqref{QCRB_simple} and twice its value, in full accord with Eq.~(6) of Ref.~\cite{Miao_PRL_119_050801_2017}. The doubling occurs if $\hbar|\Im K|$ approaches $S_{FF}$, which is its maximal value allowed by the Fluctuation-Dissipation Theorem. It is well known that in optical resonators this happens at the optical resonance point $\Omega=\delta$ in the resolved-sideband regime $\gamma\ll\delta$, where $\delta$ is the optical detuning and $\gamma$ is the optical half-bandwidth, which fully corresponds to Fig.~4 of Ref.~\cite{Miao_PRL_119_050801_2017}.

\section{Optical interferometric position meter with an auxiliary spin system}\label{sec:spin_system}

An important practical example of a stationary optical position meter with $\sigma \ne 0$ is the negative-mass-reference-frame scheme of Refs.~\cite{Hammerer_PRL_102_020501_2009, Moeller_Nature_547_191_2017} mentioned previously (see Refs.~\cite{Tsang_PRX_2_031016_2012,Polzik_AnnPhys_527_A15_2014} for a more general discussion). In this scheme the same light sequentially probes a mechanical position $x$ (using an ordinary interferometric position meter) and the collective spin of an atomic ensemble. The second interaction can be viewed as a measurement of the position $x_{\rm spin}$ of an effective harmonic oscillator with negative effective mass and an eigenfrequency equal to the Larmor precession frequency of the spin ensemble~\cite{Julsgaard_Nature_413_400_2001}. It is interesting that this scheme can be considered both as a QND measurement of the autocommuting variable $x+x_{\rm spin}$ as well as a back-action-evading measurement with frequency-dependent $S_{xF}$.
The scheme can also be implemented with two atomic spin oscillators that have effective masses with opposite signs, realizing an ac magnetometer~\cite{Wasilewski_PRL_104_133601_2010}.

The quantum noise spectral densities for this scheme, in the particular case of a broadband and resonance-tuned interferometer, are calculated in Appendix~\ref{app:neg_mass},
\begin{subequations}\label{S_spin}
  \begin{gather}
    S_{xx}(\Omega) = \frac{\hbar}{4\theta_I}
      \bigl\{1 + 4\theta_S^2|\chi_S(\Omega)|^2 + 4\theta_S|\Im\chi_S(\Omega)|\bigr\} \,, \\
    S_{FF}(\Omega) = \hbar\theta_I \,, \\
    S_{xF}(\Omega) = \hbar\theta_S \chi_S(\Omega) \,,
  \end{gather}
\end{subequations}
where $\theta_I$ and $\theta_S$ are the non-negative coupling factors for, respectively, the mechanical probe and the spin system, and $\chi_S$ is the effective susceptibility of the negative-mass system, see details in Ref.~\cite{18a1KhPo}. It is easy to see that these spectral densities satisfy and saturate the uncertainty relation~\eqref{eq:Heisenberg-general} (with $K=0$), with the term $\sigma$ originating from the dissipation in the spin system:
\begin{equation}
  S_{xx}(\Omega)S_{FF}(\Omega)-|S_{xF}(\Omega)|^{2}
  = \hbar^2\theta_S|\Im\chi_S(\Omega)| + \frac{\hbar^{2}}{4} \,.
\end{equation}

According to the reasoning of Refs.~\cite{Hammerer_PRL_102_020501_2009, Moeller_Nature_547_191_2017}, in order to cancel the back action, evolution of the negative-mass oscillator should mirror, with opposite sign, the evolution of the mechanical probe,
\begin{equation}\label{spin_subopt}
  \theta_S\chi_S(\Omega) = -\theta_I\chi(\Omega)\,.
\end{equation}
Substitution of the spectral densities~\eqref{S_spin} into Eq.~\eqref{S_sum} with account of this gives that
\begin{equation}\label{S_sum_spin}
  S_{\rm sum}(\Omega)
  = \hbar|\Im\chi^{-1}(\Omega)| + \frac{\hbar^2}{4|\chi(\Omega)|^2S_{FF}(\Omega)} \,.
\end{equation}
This spectral density is plotted in Fig.~\ref{fig:plot_S_sum}. It is easy to see that it is noticeably larger than the optimal one~\eqref{S_sum_min0}. The reason for this is the different criteria of optimization in these two cases: the condition~\eqref{spin_subopt} cancels the influence of the back-action noise, while the condition~\eqref{S_xF_opt} minimizes the sum quantum noise, which includes also the negative-mass system's thermal noise proportional to $|\Im\chi_S|$.

Substitution of the spectral densities~\eqref{S_spin} into Eq.~\eqref{S_xF_opt} gives the following requirement for the negative-mass system's effective susceptibility:
\begin{equation}\label{S_xF_opt_spin}
  \theta_S\chi_S(\Omega) = \begin{cases}
      -\theta_I\Re\chi(\Omega) \,,\
        \text{if}\ \theta_I|\Im\chi(\Omega)| < \dfrac{1}{2}\,,\\[1ex]
      -\theta_I\chi(\Omega) + \dfrac{i\hbar}{2}\,\sign[\Im\chi(\Omega)]\,, \\
        \qquad\qquad
        \text{if}\ \theta_I|\Im\chi(\Omega)| \ge \dfrac{1}{2}\,.
    \end{cases}
\end{equation}
That is, the real part of $\theta_S\chi_S(\Omega)$ indeed has to mirror the real part of $\theta_I\chi$, but the imaginary part has to be smaller than the imaginary part of $\theta_I\chi$ (and equal to zero in the ``weak back action'' case).

\section{Physical origin of the DQL}\label{sec:DQL}

It follows from the results of Sec.~\ref{sec:Sens-limit}, that the DQL is inherent in linear stationary systems. In order to reveal its physical origin, it is instructive to return to the general non-stationary case and to the time-domain picture.

Consider again the structure of the output signal of the linear position meter, see Eq.~(\ref{tilde_F}). Here $\tilde{F}$, being the {\it meter output}, is a classical observable. Therefore, its autocommutator vanishes,
\begin{equation}
  [\tilde{F}(t),\tilde{F}(t')] = 0\,.
\end{equation}
This means that the autocommutators of the meter sum noise $\hat{F}_{\rm sum}$
and of the probe thermal noise $\hat{F}_T$ must cancel each other:
\begin{equation}
  [\hat{F}_{{\rm sum}}(t),\hat{F}_{{\rm sum}}(t')]
  = -[\hat{F}_{T}(t),\hat{F}_{T}(t')] = -C_{TT}(t,t') \,.
\end{equation}
Direct calculation using Eqs.\,(\ref{probe_Kubo_F}, \ref{meter_Kubo}, \ref{F_sum_t_raw})
gives that this is indeed the case.
It is easy to show then, using the same logic as in the derivation of Eq.~\eqref{S_probe} from Eqs.~(\ref{probe_Kubo_F}, \ref{probe_td}), that in the stationary case, the spectral density of $\hat{F}_{{\rm sum}}$ cannot be smaller than the DQL~(\ref{DQL}).

It follows from this consideration that the DQL originates from the non-autocommutativity
of the probe thermal noise, similar to the SQL, which originates from
non-autocommutativity of the probe position operator. In fact, it can
be seen from Eq.~(\ref{probe_x_eq}) that instead of just the
classical ($c$-number) signal force $F_{{\rm sig}}$, its combination with
the operator-valued thermal force $\hat{F}_{T}$ is measured. The non-autocommutativity
of the latter ``contaminates'' this combination, thereby preventing
its exact continuous measurement.

This similarity with the SQL suggests a way to overcoming the DQL,
namely the use of non-stationary measurements which only give information
about some auto-commuting part of the thermal force. As an example,
consider the case of a near-resonance force acting on a harmonic oscillator.
Such a force can be decomposed into a superposition of its cosine
and sine quadratures, which excite, respectively, sine and cosine
quadratures of the probe oscillator position. Therefore a non-stationary
measurement sensitive to only one of the position quadratures will
only give information about the corresponding force quadrature, evading
the problem with the non-autocommutativity of the thermal force. It
is easy to show that the early proposals aimed at beating the SQL,
by means of measuring only one probe quadrature~\cite{78a1eBrKhVo, Thorne1978, Caves_RMP_52_341_1980, Caves_PRD_26_1817_1982, Buchmann_PRL_117_030801_2016}, allow the overcoming of not only the SQL, but also the DQL.

\section{Conclusions and outlook}\label{sec:conclusion}

In this work we have presented a general analysis of the ultimate quantum sensitivity limits that pertain to stationary force sensing. We employed a linear response formalism permitting a generic analysis in terms of the probe system susceptibility function and meter system correlation functions. Our analysis  simultaneously includes the effects of dissipative probe dynamics, dynamic back action, and finite probing strength.
This approach allowed us to derive the general quantum limits~(\ref{S_sum_min0},~\ref{S_sum_simple_0}) and~(\ref{S_sum_min},~\ref{S_sum_min_simple}) for the force sensitivity (for fixed $S_{\F\F}$ and $S_{FF}$, respectively), and the requirements on the meter noise spectral densities for attaining these limits. In particular, we elucidated the transition between a force sensor being limited by the QCRB versus the DQL. We showed here that the imaginary meter cross-correlations $\Im S_{xF}$, originating from dissipation in the meter, can be beneficial for achieving the DQL. We exemplified this using a negative-mass spin oscillator that allows the generation of such meter field correlations. We also revealed the physical origin of the DQL, namely, the non-autocommutativity of the probe thermal noise force, which must necessarily be matched by a corresponding non-autocommutativity of the meter sum quantum noise force.

In addition to establishing the fundamental quantum sensing limits, our generic results may serve as useful tools in the design and optimization of quantum-limited sensors. The QCRB (in the form of the {\it shot noise}) is already one of the main sensitivity limitations in the large-scale GW laser interferometers. At the same time, the DQL could be an important limitation in table-top optomechanical experiments at frequencies in the vicinity of the mechanical resonance, where $\Im\chi^{-1}$ could be comparable with $|\chi^{-1}|$ and therefore the DQL approaches the SQL, see, \eg, Refs.~\cite{Kampel_PRX_7_021008_2017,Mason_NPhys_15_745_2019}; the same consideration applies to magnetometers based on atomic spin oscillators~\cite{Wasilewski_PRL_104_133601_2010}.

In the present work, we have considered the local optimization of the quantum noise at a given signal (Fourier) frequency, establishing in this way the ultimate sensitivity limits. At the same time, in the practical design of any force sensor, the sensing bandwidth is a crucial consideration. A pertinent question is therefore whether the meter correlations required to achieve quantum-limited performance can be feasibly engineered over a bandwidth suitable for the sensing application at hand. This aspect will be discussed in our subsequent publication~\cite{Zeuthen-etal_InPrep}.

\begin{acknowledgments}

The authors would like to thank E.\,S.\ Polzik and O.\ Sandberg for commenting on the manuscript.

This work has been supported by the European Research Council Advanced grant QUANTUM-N and the Villum Foundation. 
The work of F.\,K.\ was supported by the Russian Foundation for Basic Research grant 19-29-11003.

\end{acknowledgments}

\appendix

\section{Derivation of Eqs.~\eqref{eq:Heisenberg-general} and \eqref{S_probe}}\label{app:meter-uncert}

On account of Eqs.~(\ref{probe_Kubo_F},~\ref{meter_Kubo}) and the stationarity conditions~(\ref{stat_B},~\ref{stat_chi}), Eqs.~(\ref{probe_td},~\ref{Heisenberg_td}) can be stated in the Fourier domain as
\begin{subequations}\label{Omega_int}
  \begin{gather}
    \intinfty|Q(\Omega)|^2\bigl\{S_{TT}(\Omega) - \hbar\Im\chi^{-1}(\Omega)\bigr\} d\Omega
      \ge 0 \,, \\
    \sum_{p,q\in\{x,F\}}\int_{-\infty}^{\infty} Q_p^*(\Omega)Q_q(\Omega)
      \bigl\{S_{pq}(\Omega) + C_{pq}(\Omega)/2\bigr\} d\Omega \ge 0 \,,
    \label{Heisenberg_Omega}
  \end{gather}
\end{subequations}
for arbitrary $Q(\Omega)$, $Q_{p,q}(\Omega)$, where the spectra of the commutators $C_{pq}$ are defined by
\begin{equation}\label{stat_C}
  C_{pq}(t,t') = C_{pq}(t-t',0)
  = \intinfty C_{pq}(\Omega)e^{-i\Omega(t-t')}\,\frac{d\Omega}{2\pi} \,,
\end{equation}
and are equal to
\begin{subequations}\label{meter_Kubo_Omega}
  \begin{gather}
    C_{xx}(\Omega) = 0 \,, \\
    C_{FF}(\Omega) = -2\hbar\Im K(\Omega) \,, \\
    C_{xF}(\Omega) = -i\hbar\,.
  \end{gather}
\end{subequations}
The following inequalities, which are local in $\Omega$, are necessary and sufficient conditions for Eqs.~\eqref{Omega_int}:
\begin{subequations}
  \begin{gather}
    S_{TT}(\Omega) - \hbar\Im\chi^{-1}(\Omega) \ge 0 \,, \label{probe_plus} \\
    \sum_{p,q\in\{x,F\}}Q_p^*Q_q\bigl
      \{S_{pq}(\Omega) + C_{pq}(\Omega)/2\bigr\} \ge 0 \,,
      \label{Heisenberg_Omega_loc}
  \end{gather}
\end{subequations}
where now $Q_{p,q}$ are arbitrary complex numbers. Due to Eqs.~\eqref{meter_Kubo_Omega} and the symmetry properties,
\begin{subequations}
  \begin{gather}
    S_{pq}(\Omega) = S_{qp}^*(\Omega)\,, \\
    C_{pq}(\Omega) = C_{qp}^*(\Omega)\,,
  \end{gather}
\end{subequations}
Eq.~\eqref{Heisenberg_Omega_loc} can be reduced to
\begin{multline}\label{meter_plus}
  S_{xx}(\Omega)\bigl\{S_{FF}(\Omega) - \hbar\Im K(\Omega)\bigr\} \\
    \ge |S_{xF}(\Omega)|^2 - \hbar\Im S_{xF}(\Omega) + \frac{\hbar^2}{4} \,.
\end{multline}
Substituting $\Omega\to-\Omega$ in Eqs.~(\ref{probe_plus},~\ref{meter_plus}) and taking into account the symmetry properties
\begin{subequations}
  \begin{gather}
    S_{qp}(-\Omega) = S_{qp}^*(\Omega)\,, \\
    K(-\Omega) = K^*(\Omega)\,,
  \end{gather}
\end{subequations}
we obtain
\begin{subequations}
  \begin{equation}
    S_{TT}(\Omega) + \hbar\Im\chi^{-1}(\Omega) \ge 0 \,, \label{probe_minus} \\
  \end{equation}
  \begin{multline}
    S_{xx}(\Omega)\bigl\{S_{FF}(\Omega) + \hbar\Im K(\Omega)\bigr\} \\
      \ge |S_{xF}(\Omega)|^2 + \hbar\Im S_{xF}(\Omega) + \frac{\hbar^2}{4} \,.
      \label{meter_minus}
  \end{multline}
\end{subequations}
Finally, the combination of Eqs.~(\ref{probe_plus},~\ref{probe_minus}) yields Eq.~\eqref{S_probe}, whereas that of Eqs.~(\ref{meter_plus},~\ref{meter_minus}) yields Eq.~\eqref{eq:Heisenberg-general}.

\section{Derivation of Eqs.~(\ref{S_sum_min}) and (\ref{S_sum_min_simple})}\label{app:opt}

\paragraph{Notations.}

For brevity, we suppress in this Appendix the explicit frequency dependence of all variables, set $\hbar=1$, and introduce the following notations:
\begin{subequations}
  \begin{equation}
    D = \chi^{-1} \,, \quad D_K = \chi_K^{-1} \,,
  \end{equation}
  and for any quantity $Q$,
  \begin{equation}
    Q' = \Re Q \,, \quad Q'' = \Im Q \,.
  \end{equation}
\end{subequations}

In this notation, Eqs.~(\ref{eq:Heisenberg-general},~\ref{S_sum}) have the following form:
\begin{subequations}
  \begin{gather}
    S_{xx}S_{FF} - S_{xF}'^2 - (K''S_{xx} - \sigma)^2 = |\sigma| + \frac{1}{4} \,,
      \label{Heis_1} \\
    S_{\rm sum} = (|D_K|^2 - 2D_K''K'')S_{xx} + 2D_K'S_{xF}'
      + 2D_K''\sigma + S_{FF} \,,
  \end{gather}
\end{subequations}
or
\begin{subequations}
  \begin{gather}
    X^2 + Y^2 = \frac{R^2}{4K''^4} \,, \label{X2Y2} \\
    S_{\rm sum} = AX + BY + C \,,
  \end{gather}
\end{subequations}
where
\begin{subequations}
  \begin{gather}
    X = S_{xx}  - \frac{\sigma}{K''} - \frac{S_{FF}}{2K''^2} \,,
    \quad Y = \frac{S_{xF}'}{K''} \,, \\
    R 
    = \sqrt{(S_{FF} - K''\sign\sigma)\{S_{FF} + K''(4|\sigma|+1)\sign\sigma\}} \,, \\
    A = |D_K|^2 - 2D_K''K'' = D_{K}'^2+D''^2 - K''^2 \,, \\
    B = 2D_K'K'' \,, \\
    C = \frac{1}{2K''^2}\bigl[(|D_K|^2 - 2D''K'')S_{FF} + 2|D_K|^2K''\sigma\bigr] \,.
  \end{gather}
\end{subequations}

\paragraph{Optimization in $S_{xx}$ and $\text{Re} S_{xF}$.}

It follows from Eq.~\eqref{X2Y2} that
\begin{equation}
  X = \frac{R}{2K''^2}\cos\theta \,, \quad Y = \frac{R}{2K''^2}\sin\theta \,,
\end{equation}
where
\begin{equation}
  \theta = \arctan\frac{Y}{X} \,.
\end{equation}
Therefore,
\begin{equation}
  S_{\rm sum} = \frac{R}{2K''^2}(A\cos\theta + B\sin\theta) + C \,.
\end{equation}

The minimum of $S_{\rm sum}$ in $\theta$ is provided by
\begin{equation}
  \cos\theta = -\frac{A}{\sqrt{A^2+B^2}} \,, \quad
  \sin\theta = -\frac{B}{\sqrt{A^2+B^2}} \,,
\end{equation}
which is equivalent to
\begin{subequations}\label{S_subopt}
  \begin{gather}
    S_{xx} = \frac{1}{2K''^2}
      \biggl(S_{FF} - \frac{|D_K|^2 - 2D_K''K''}{\mathcal{D}^2}\,R\biggr)
       + \frac{\sigma}{K''} \,, \\
    S_{xF}' = -\frac{D_K'}{\mathcal{D}^2}\,R \,,
  \end{gather}
\end{subequations}
where
\begin{multline}
  \mathcal{D}^4 = A^2 + B^2 = |D_K|^2(|D_K|^2 - 4D''K'') \\
  = 4D''^2(S_{\rm thr}^2-K''^2) \,,
\end{multline}
and
\begin{equation}
  S_{\rm thr} = \frac{|D_K|^2 - 2D''K''}{2|D''|} = \frac{D_{K}'^2+D''^2+K''^2}{2|D''|}\,.
\end{equation}
The minimum value is
\begin{multline}\label{S_sum_subopt}
  S_{\rm sum} = C - \frac{R}{2K''^2}\sqrt{A^2+B^2} \\
  = \frac{1}{2K''^2}\bigl\{(|D_K|^2 - 2D''K'')S_{FF} + 2|D_K|^2K''\sigma
    - \mathcal{D}^2R\bigr\} \,.
\end{multline}

\paragraph{$\sigma=0$.}

In this case,
\begin{multline}\label{S_subopt_0}
  S_{\rm sum} = \frac{|D''|}{K''^2}
    \Bigl\{S_{\rm thr}S_{FF} - \sqrt{(S_{\rm thr}^2-K''^2)(S_{FF}^2 - K''^2)}\Bigr\} \\
  = \frac{|D''|(S_{\rm thr}^2+S_{FF}^2 - K''^2)}
      {S_{\rm thr}S_{FF} + \sqrt{(S_{\rm thr}^2-K''^2)(S_{FF}^2 - K''^2)}} \,;
\end{multline}
(note that $S_{FF},S_{\rm thr}\geq |K''|$).
The minimum of Eq.~\eqref{S_subopt_0} in $S_{FF}$ is provided by
\begin{equation}
  S_{FF} = S_{\rm thr}
\end{equation}
and is equal to the DQL,
\begin{equation}
  S_{\rm sum} = |D''| \,.
\end{equation}

\paragraph{Minimum in $\sigma$.}

Minimizing $S_{\rm sum}(\Omega)$, Eq.~\eqref{S_sum_subopt}, with respect to $|\sigma|$, we find that a non-zero minimum point $|\sigma|>0$ must obey
\begin{equation}
|\sigma| = |D''|\frac{-S_{FF}\sign(\sigma D'') - S_{\rm thr}}{|D_K|^2}\,,
\end{equation}
and hence exists if $S_{FF}>S_{\rm thr}$ with $\sign\sigma = -\sign D''$. The corresponding minimum value of $S_{\rm sum}$ is the DQL,
\begin{equation}
  S_{\rm sum} = |D''| \,.
\end{equation}

If $S_{FF}\leq S_{\rm thr}$, the minimum of Eq.~\eqref{S_sum_subopt} occurs at $\sigma=0$ and is given by Eq.~\eqref{S_subopt_0}.

\section{Negative mass reference frame system}\label{app:neg_mass}

In this Appendix, we suppress for brevity the explicit frequency dependence of all variables.

Input/output relations for the optical interferometric position meter, for the particular case of the broadband resonance-tuned interferometer are the following, see, \eg, Ref.~\cite{16a1DaKh}:
\begin{subequations}\label{eq:I-IO}
  \begin{gather}
    \hat{{\rm b}}_I^c = \hat{{\rm a}}_I^c \,, \\
    \hat{{\rm b}}_I^s = \hat{{\rm a}}_I^s  + \sqrt{\frac{2\theta_I}{\hbar}}\,\hat{x}\,, \\
    \hat{F}_{\rm fl} = \sqrt{2\hbar\theta_I}\,\hat{{\rm a}}_I^c \,,
  \end{gather}
\end{subequations}
where $\hat{{\rm a}}_I^{c,s}$ and $\hat{{\rm b}}_I^{c,s}$ are the cosine and sine quadratures of the incident and output light.
The corresponding relations for the atomic spin system are~\cite{18a1KhPo}
\begin{subequations}\label{eq:S-IO}
  \begin{gather}
    \hat{{\rm b}}_S^c = \hat{{\rm a}}_S^c \,, \\
    \hat{{\rm b}}_S^s = \hat{{\rm a}}_S^s + 2\theta_S\chi_S\hat{{\rm a}}_S^c
      + \sqrt{2\theta_S}\,\chi_S\hat{F}_S \,.
  \end{gather}
\end{subequations}

Based on Eqs.~(\ref{eq:I-IO},~\ref{eq:S-IO}), it is easy to show that independently of whether the light probes first the interferometer and after that the atomic spin system or vice versa, the input/output relations for the combined system are
\begin{subequations}
  \begin{gather}
    \hat{{\rm b}}^s = \hat{{\rm a}}^s + 2\theta_S\chi_S \hat{{\rm a}}^c
      + \sqrt{2\theta_S}\,\chi_S\hat{F}_S + \sqrt{\frac{2\theta_I}{\hbar}}\,\hat{x}
      \propto \hat{x}_{\rm fl} + \hat{x} \,,  \\
    \hat{F}_{\rm fl} = \sqrt{2\hbar\theta_I}\,\hat{{\rm a}}^c \label{B:F_fl} \,,
  \end{gather}
\end{subequations}
where $\hat{{\rm a}}^{c,s}$ and $\hat{{\rm b}}^{c,s}$ are the cosine and sine quadratures of the incident and output light, respectively, for the combined system, and
\begin{equation}
  \hat{x}_{\rm fl} = \sqrt{\frac{\hbar}{2\theta_I}}\bigl(
      \hat{{\rm a}}^s + 2\theta\chi_S \hat{{\rm a}}^c + \sqrt{2\theta_S}\,\chi_S\hat{F}_S
    \bigr) .
\end{equation}

We assume that the incident field is in the ground state, that is the spectral densities of $\hat{{\rm a}}^{c,s}$ are equal to $1/2$. In this case, the spectral densities of $\hat{x}_{\rm fl}$, $\hat{F}_{\rm fl}$, and their cross-spectral density are given by Eqs.~\eqref{S_spin}.

\end{document}